\documentclass[journal]{article}
\usepackage{amsfonts}

\begin{document}

\title{Space~Efficient~Secret~Sharing: A~Recursive~Approach}

\author{Abhishek~Parakh and~Subhash~Kak%
\thanks{A. Parakh and S. Kak are with the Computer Science
Department, Oklahoma State University, Stillwater, OK, 74078 USA
e-mail: (see http://cs.okstate.edu/\~{}parakh and http://cs.okstate.edu/\~{}subhashk).}}

\maketitle

\begin{abstract}
This paper presents a recursive secret sharing technique that distributes $k-1$ secrets of length $b$ each into $n$ shares such that each share is effectively of length $\frac{n}{k-1}\cdot b$ and any $k$ pieces suffice for reconstructing all the $k-1$ secrets. Since $\frac{n}{k-1}$ is near the optimal factor of $n/k$, and can be chosen to be close to 1, the proposed technique is space efficient. Furthermore, each share is information theoretically secure, i.e. it does not depend on any unproven assumption of computational intractability. Such a recursive technique has potential applications in secure and reliable storage of information on the Web and in sensor networks.
\end{abstract}

\section{Introduction}
Conventional secret sharing schemes, although information theoretically secure, are space inefficient. Thus $k$-out-of-$n$ secret sharing techniques have a "blow-up" factor of $n$, i.e. sharing a secret of size $b$ requires a total storage space of size $b\cdot n$.


In order to improve space efficiency, computational secret sharing techniques have been developed \cite{ref2,ref3,ref4, ref10} in which a symmetric key is used to encrypt the original secret and the key is split into shares using conventional methods of secret sharing. The encrypted secret is divided into pieces to which redundancy is added by the use of block error correction techniques \cite{ref7,ref8,ref9}. This leads to a $n$-fold increase in key size, pieces of which have to be stored with every share of the encrypted secret, hence becoming a overhead. Moreover, this reduction in storage is achieved by relaxing the security requirements, since the computational security is weaker than information theoretic security \cite{ref9}.

Consequently, we ask whether it is possible to improve space efficiency of secret sharing techniques while maintaining information theoretic security. This question is answered, herein, in the affirmative by proposing a secret sharing scheme that encodes $k-1$ secrets in $n$ shares compared to conventional methods of encoding 1 secret in $n$ shares, thus increasing space efficiency and maintaining information theoretic security.

In an earlier paper \cite{ref5}, a 2-out-of-2 ($k=2$ and $n=2$) recursive scheme for secret sharing was proposed. In this method, if $k$ secrets are chosen such that they double in size, then all of the smaller secrets can be recursively stored in the shares of larger secrets, so that two shares of size $2^m$ can encode $2^{m+1}-1$ bits of information. For example, if we are to share 3 secrets $s_1=1$, $s_2=01$, and $s_3=1011$, then the two shares for $s_1$ would be $D_{{s_1}1}=0$ and $D_{{s_1}2}=1$; where exclusive-OR operation is used for secret reconstruction. The shares of $s_1$ can be used to create two shares of $s_2$ as follows: $D_{{s_2}1}=D_{{s_1}1}0=00$ and $D_{{s_2}2}=0D_{{s_1}2}=01$. Here $D_{{s_1}1}0$ denotes concatenation of share 1 of secret $s_1$ with 0; and $0D_{{s_1}2}$ denotes concatenation of 0 with share 2 of secret $s_1$, and so on. Similarly, we can recursively use the shares of $s_2$ to create the shares of $s_3$: $D_{{s_3}1}=D_{{s_2}1}10=0010$ and $D_{{s_3}2}=10D_{{s_2}2}=1001$. As a result, the final two shares for all the three secrets are 0010 and 1001. These shares have recursively encoded within themselves the shares of smaller secrets. Consequently, 8 bits of shares have encoded 7 bits of secrets. This is in comparison with conventional methods that would require 14 bits of shares. Note that the example presented here is slightly modified to improve the security of the scheme as compared to that presented in \cite{ref5}. This idea of recursive secret sharing has been extended to 2-out-of-$n$ secret sharing in \cite{ref6}, requiring a $n$-fold increase in secret sizes at each step.

The schemes presented in \cite{ref5,ref6} that build upon the requirement of $n$-fold increase in secret sizes at each step, only reduce the conventional blow up factor of $n$ to $n-1$.

In this paper we relax the requirement of $n$-fold increase in secret sizes at each step and present a general recursive $k$-out-of-$n$ scheme that increases the efficiency of secret sharing to nearly 100\% with no restrictions on increase in secret sizes, while maintaining information theoretic security. A result of this relaxation on secret sizes is that we are able to reduce the blow factor from $n$ to $\frac{n}{k-1}$. In more precise terms, we present a scheme that can store up to $k-1$ secrets of size $b$ each using $n$ shares such that each share is effectively of size $(\frac{n}{k-1})\cdot b$. This is in comparison with conventional secret sharing schemes that store just one secret of size $b$ in $n$ shares, requiring a total storage of size $n\cdot (k-1)\cdot b$ for $k-1$ secrets. Hence our scheme is near optimal with a blow up factor of $\frac{n}{k-1}$ which can be chosen to be close to 1. Further the scheme is information theoretically secure and does not depend on any computational assumptions.

The proposed recursive secret sharing scheme has applications in distributed online storage of information discussed in \cite{ref7,ref8}. However, Rabin \cite{ref7} discarded the possibility of using secret sharing schemes because of the  $n$-fold increase in storage space required by conventional implementations of secret sharing techniques. Since our scheme provides a near optimal way to encode data into small shares with no information leakage, it becomes an ideal candidate for use in secure online data storage. Such secure data storage scheme is an example of implicit data security since the word \textit{implicit} conveys the idea that there is no explicit encryption of data and no encryption keys are used. Data is so divided that each piece is implicitly secure in itself and only reveals information when $k$ or more of the pieces are brought together.

\section{Space efficient secret sharing}

Below, we define the terms used in this paper for a $(k,n)$ secret sharing scheme, where any $k$ out of $n$ shares suffice to reconstruct the secret. Since we are working with information theoretically secure schemes $b$ denotes the size of secret/s as well as the size of share.\\

\noindent \textbf{Definition 1.} \textit{Blow-up factor (secret sharing)}\\\\
    $=\frac{\textrm{total size of shares}}{\textrm{total size of secret/s encoded by the shares}}$ \\\\
    $=\frac{\textrm{number of shares $\times$ size of a share}}{\textrm{total size of original secret/s}}$
\\\\
\noindent \textbf{Definition 2.} \textit{Blow-up factor (conventional secret sharing)}\\\\
    $=\frac{\textrm{number of shares $\times$ size of a share}}{\textrm{total size of original secret/s}}$
    $=\frac{n\times b}{b}=n$
\\\\
\noindent \textbf{Definition 3.} \textit{Space optimal secret sharing scheme}: A secret sharing scheme with a blow up factor of $\frac{n}{k}$, where $2\leq k\leq n$. This is because for an space optimal secret sharing scheme\\\\
    $\frac{\textrm{number of shares $\times$ size of a share}}{\textrm{total size of original secret/s}}$
    $=\frac{n\times b}{k\times b}=\frac{n}{k}$
\\\\
\noindent \textbf{Definition 4.} \textit{Space efficient secret sharing scheme}: A secret sharing scheme that approaches the optimal blow up factor of $\frac{n}{k}$.\\

The proposed scheme recursively builds upon Shamir's secret sharing scheme. Since it is known that Shamir's scheme is information theoretically secure, the proposed scheme is also similarly secure. However, Shamir's scheme generates $n$ shares for every single secret, giving rise to a $n$-fold increase in storage space. Whereas, we share $k-1$ secrets using $n$ shares resulting in only $\frac{n}{k-1}$-fold increase in storage space, which is near optimal.

We first briefly review the scheme presented by Shamir \cite{ref1}. Note that Shamir only encoded one secret $s$ in $n$ shares as discussed in Algorithm 1.\\

\noindent \verb"Algorithm 1(Shamir's secret sharing scheme)"
\begin{enumerate}
\item Choose a prime $p$, $p>max(s,n)$, where $s\in \mathbb{Z}_p$ is the secret.
\item Choose $k-1$ random numbers $a_1$, $a_2$, ..., $a_{k-1}$, uniformly and independently, from the field $\mathbb{Z}_p$.
\item Using $a_i$, $1\leq i\leq (k-1)$ and secret $s$, generate polynomial $p(x)$ of degree $k-1$,\\ $p(x)=s+a_1x+a_2x^2+...+a_{k-1}x^{k-1}$  (mod p).
\item Sample $p(x)$ at $n$ points $D_i=p(i)$, $1\leq i\leq n$ such that the shares are given by $(i,D_i)$.\\
\end{enumerate}

The reconstruction of the secret is performed by interpolating any $k$ points (shares) and evaluating $s=p(0)$. Algorithm 1 is known to be information theoretically secure due to the properties of interpolation, which says that $k-1$ shares do not reveal any information.

Throughout the paper, we work in a finite field $\mathbb{Z}_p$, where $p$ is a prime and $p\geq max(s_{max},n)$, where $s_{max}=max(s_i)$,$1\leq i\leq (k-1)$, and $s_1$, $s_2$, ..., $s_{k-1}$ are the secrets. The shares will be denoted as $D_{{s_i}1}$, $D_{{s_i}2}$, ...,$D_{{s_i}m}$ at the intermediate stages, where $2\leq m\leq k-2$ and $D_1$, $D_2$, ..., $D_n$ at the final stage. (Note that, as in Shamir's scheme, $D_{{s_i}m}$'s are the y-coordinates only, while the respective x-coordinates $m$'s, are implicitly known to all players.)

The intuition for the proposed scheme is as follows: We randomly and uniformly choose a number $a_1\in\mathbb{Z}_p$ and generate $1^{st}$ degree polynomial $p_1(x)=a_1x+s_1$. Then we sample $p_1(x)$ at two points $D_{{s_1}1}=p_1(1)$ and $D_{{s_1}2}=p_1(2)$, to generate two shares for $s_1$. This first step can be viewed as a direct execution of Shamir's (2, 2) secret sharing scheme. Next we use these two shares of $s_1$ to generate polynomial $p_2(x)=D_{{s_1}2}x^2+D_{{s_1}1}x+s_2$, where the coefficients are the previous two shares and the free term is the new secret. Sampling $p_2(x)$ at three points $D_{{s_2}1}=p_2(1)$, $D_{{s_2}2}=p_2(2)$, and $D_{{s_2}3}=p_2(3)$, generates three shares of $s_2$. We can now delete $D_{{s_1}1}$ and $D_{{s_1}1}$ because the new shares $D_{{s_2}1}$, $D_{{s_2}2}$, and $D_{{s_2}3}$ have the shares of $s_1$ hidden within themselves. We then use the shares of $s_2$ to create a $3^{rd}$ degree polynomial with $s_3$ as its free term and generate shares for $s_3$ by sampling the newly created polynomial at 4 points. These four points denoted as $D_{{s_3}1}$, $D_{{s_3}2}$, $D_{{s_3}3}$, and $D_{{s_3}4}$ have the shares of $s_1$, $s_2$ as well as $s_3$ and therefore $D_{{s_2}1}$, $D_{{s_2}2}$ and $D_{{s_2}3}$ can now be deleted.  The process is repeated for secrets $s_4$, $s_5$, ..., $s_{k-1}$ by creating $p_4(x)$, $p_5(x)$, ...,$p_{k-2}(x)$ and repetitive sampling and reusing of shares and deleting the older shares. At the last step, we generate a polynomial $p_{k-1}(x)=D_{{s_{k-2}}(k-1)}x^{k-1}+D_{{s_{k-2}}(k-2)}x^{k-2}+...+D_{{s_{k-2}}1}x+s_{k-1}$ and sample it at $n$ points $D_1=p_{k-1}(1)$, $D_2=p_{k-1}(2)$, ..., $D_n=p_{k-1}(n)$, such that the final shares are given by $(i,D_i)$, $1\leq i\leq n$. These final $n$ shares have recursively hidden $k-1$ secrets within themselves. Algorithm 2 illustrates the process.\\

\noindent \verb"Algorithm 2 - Dealing Phase"
\begin{enumerate}
\item \noindent Randomly and uniformly choose a number $a_1\in\mathbb{Z}_p$ and generate polynomial $p_1(x)=a_1x+s_1$.
\item Sample $p_1(x)$ at two points $D_{{s_1}1}=p_1(1)$ and $D_{{s_1}2}=p_1(2)$, which represent two shares of $s_1$.
\item Do for $2\leq i\leq (k-1)$

\begin{enumerate}
\item Generate polynomial,\\ $p_i(x)=D_{s_{i-1}i}x^i+D_{s_{i-1}(i-1)}x^{i-1}+...+D_{s_{i-1}1}x+s_i$.
\item Sample $p_i(x)$ to create new shares,

\begin{enumerate}
\item If $i<k-1$, sample at $i+1$ points:\\
    $D_{{s_i}1}=p_i(1)$\\
    $D_{{s_i}2}=p_i(2)$\\
    \vdots\\
    $D_{{s_i}(i+1)}=p_i(i+1)$.
\item If $i=k-1$, sample at $n$ points:\\
    $D_1=p_i(1)$\\
    $D_2=p_i(2)$\\
    \vdots\\
    $D_n=p_i(n)$.
\end{enumerate}

\item If $i<k-1$, delete old shares: $D_{s_{i-1}1}$, $D_{s_{i-1}2}$, ..., $D_{s_{i-1}i}$.
\end{enumerate}

\item The final $n$ shares are explicitly given by $(i,D_i)$, $1\leq i\leq n$.\\
\end{enumerate}

For a the trivial case of just one secret $s_1$ and a $(2,n)$ secret sharing, the algorithm stops at step 2, where we sample the polynomial of first degree $p_1(x)$ at $n$-points to create $n$ shares such that any 2 of the shares can reconstruct $s_1$.\\

\noindent \verb"Algorithm 2 - Reconstruction Phase"
\begin{enumerate}
\item Interpolate any $k$ shares $(i,D_i)$ to generate the polynomial of degree $k-1$,\\
$p_{k-1}(x)=D_{s_{k-2}(k-1)}x^{k-1}+D_{s_{k-2}(k-2)}x^{k-2}+...+D_{s_{k-2}1}x+s_{k-1}$\\
and evaluate $s_{k-1}=p_{k-1}(0)$.
\item Do for all $i=k-2$ down to 1

\begin{enumerate}
\item Interpolate $i+1$ shares given by $(m+1, D_{s_i (m+1)})$, $0\leq m\leq i$ obtained from coefficients of $p_{i+1}(x)$ to generate polynomial of degree $i$,\\ $p_i(x)=D_{s_{i-1}i}x^i+D_{s_{i-1}(i-1)}x^{i-1}+...+D_{s_{i-1}1}x+s_i$.
\item Evaluate $s_i=p_i(0)$.
\end{enumerate}

\end{enumerate}

As seen above the reconstruction of secrets is straightforward. Any $k$ of the players can interpolate the polynomial of degree $k-1$ such that the free term represents $s_{k-1}=p_{k-1}(0)$. Then using the $k-1$ coefficients of this polynomial as points (leaving out the free term which is $s_{k-1}$), interpolate the polynomial of degree $k-2$ to obtain $s_{k-2}$. This process is repeated until we obtain $s_1$.

Algorithm 2 has clearly been able to share $k-1$ secrets among $n$ players such that every $k$ of them can interpolate all the $k-1$ secrets. Therefore, the blow up factor has been reduced to $\frac{n}{k-1}$.

\newtheorem{theorem}{Theorem}
\begin{theorem}
Algorithm 2 generates information theoretically secure shares.
\end{theorem}

\textbf{Proof} The proof builds up on the security of Shamir's secret sharing scheme. We know that Shamir's scheme is information theoretically secure. Note that step 1-2 in Algorithm 2 may be viewed as $(2,2)$ Shamir's secret sharing scheme. As a result, the shares for $s_1$, i.e. $D_{s_1 1}$ and $D_{s_1 2}$ are information theoretically secure. In other words, given any number $r\in\mathbb{Z}_p$, $Pr(r=D_{s_1 1})=Pr(r=D_{s_1 2})=\frac{1}{p}$.

Moreover, since $a_1$ is randomly and uniformly chosen from the field, $D_{s_1 1}$ and $D_{s_1 2}$ can be viewed as random numbers. These shares are then used as random coefficients to generate a (3,3) Shamir's secret sharing scheme, with secret $s_2$ as the free term. The shares of $s_2$ from (3,3) Shamir's scheme encode the shares of $s_1$ within themselves and can be used as coefficients to create a (4,4) Shamir's scheme with $s_3$ as the free term. Now, the shares $s_3$ from the (4,4) Shamir's scheme encode the shares of $s_1$ and $s_2$ in themselves. We then use these new shares to generate a (5,5) Shamir's with $s_4$ as the free term and so on. This process is repeated until we reach a $(k-1,k-1)$ Shamir's scheme and have encoded $k-2$ secrets.

At this point, we use the shares of $s_{k-2}$ as coefficients to generate a $(k-1)^{th}$ degree polynomial $p_{k-1}(x)$ with $s_{k-1}$ as the free term. We then sample $p_{k-1}(x)$ at $n$-points, i.e. a $(k,n)$ Shamir's scheme. Note that these final $n$-points (shares) encode the shares of $k-1$ secrets within themselves.

Our algorithm is recursive and given that the (2,2) Shamir's scheme is secure, Algorithm 2 generates information theoretically secure shares.\hspace{23mm}$\square$\\

\textbf{Example.} Let $s_1=17$, $s_2=28$, $s_3=5$, and $s_4=12$ be four secrets that are to be shared between 7 players such that any 5 of them can reconstruct all the 4 secrets. Let prime $p=31$.\\

\noindent \textit{Dealing phase}.
\begin{enumerate}
\item Randomly and uniformly choose a number $a_1\in\mathbb{Z}_p$. Let $a_1=22$. Generate polynomial, $p_1(x)=a_1x+s_1=22x+17$ (mod 31).
\item Sample $p_1(x)$ at two points to generate two shares of secret $s_1$, i.e. $D_{s_1 1}=p_1(1)=8$ and $D_{s_12}=p_1(2)=30$.
\item Generate polynomial $p_2(x)=D_{s_12}x^2+D_{s_11}x+s_2=30x^2+8x+28$.
\item Sample $p_2(x)$ at 3 points to generate three shares of $s_2$, i.e. $D_{s_21}=p_2(1)=4$, $D_{s_22}=p_2(2)=9$, and $D_{s_23}=p_2(3)=12$.
\item Delete $D_{s_1 1}$ and $D_{s_12}$.
\item Generate polynomial $p_3(x)=D_{s_23}x^3+D_{s_22}x^2+D_{s_21}x+s_3=12x^3+9x^2+4x+5$.
\item Sample $p_3(x)$ at 4 points to generate four shares of $s_3$, i.e. $D_{s_31}=p_3(1)=30$, $D_{s_32}=p_3(2)=21$, $D_{s_33}=p_3(3)=19$, and $D_{s_34}=p_3(4)=3$.
\item Delete $D_{s_21}$, $D_{s_22}$, and $D_{s_23}$.
\item Generate polynomial\\
    $p_4(x)=D_{s_34}x^4+D_{s_33}x^3+D_{s_32}x^2+D_{s_31}x+s_4=3x^4+19x^3+21x^2+30x+12$.
\item Sample $p_4(x)$ at 7 points, which represents the final 7 shares. Hence,
    $D_1=p_4(1)=23$, $D_2=p_4(2)=15$, $D_3=p_4(3)=24$, $D_4=p_4(4)=3$, $D_5=p_4(5)=8$, $D_6=p_4(6)=12$, and $D_7=p_4(7)=29$.
\item Delete $D_{s_31}$, $D_{s_32}$, $D_{s_33}$, and $D_{s_34}$.
\end{enumerate}

The final seven shares are given by $(1,D_1)=(1,23)$; $(2,D_2)=(2,15)$; $(3,D_3)=(3,24)$; $(4,D_4)=(4,3)$; $(5,D_5)=(5,8)$; $(6,D_6)=(6,12)$; and $(7,D_7)=(7,29)$.\\

\noindent \textit{Reconstruction phase}.\\
All the four secrets can be reconstructed using any 5 out of 7 final shares.

Using 5 shares, say $(1,23)$, $(3,24)$, $(4,3)$, $(5,8)$, and $(7,29)$, we can interpolate the $4^{th}$ degree polynomial $p_4(x)=3x^4+19x^3+21x^2+30x+12$ (mod 31), thus retrieving secret $s_4$ (the free term of the polynomial) by evaluating $s_4=p_4(0)$.

Then extracting the coefficients of $p_4(x)$ and using them as y-coordinates of points $x$=1, 2, 3, and 4, i.e. $(1,30)$, $(2,21)$, $(3,19)$, and $(4,3)$ we can regenerate the $3^{rd}$ degree polynomial $p_3(x)=12x^3+9x^2+4x+5$ by interpolation and retrieve the $s_3$ as the free term, $s_3=p_3(0)$.

The coefficients of $p_3(x)$ are then used as points $(1,4)$, $(2,9)$, $(3,13)$ to interpolate $p_2(x)=30x^2+8x+28$ and reconstruct $s_2=p_2(0)$.

The coefficients of $p_2(x)$ are used as $(1,8)$ and $(2,30)$ to interpolate $p_1(x)=22x+17$ and reconstruct $s_1=p_1(0)$.

The algorithm simulates a Last In First Out (LIFO) data structure.

\section{Conclusions}
We have presented a recursive scheme that distributes $k-1$ secrets amongst $n$ individuals. The scheme is general and it places no restriction on the secret size. Since this method builds upon Shamir's secret sharing scheme, it is information theoretically secure. It has a blow up factor of $\frac{n}{k-1}$ which is near the optimal blow up factor of $\frac{n}{k}$ and represents a significant improvement over conventional secret sharing schemes.

The proposed scheme has applications in secure distributed storage of information on the Web and in sensor networks.

\bibliographystyle{IEEEtran}
\bibliography{IEEEabrv,SpaceEfficientRecursive}

\end{document}